\documentclass[sort,journal]{IEEEtran}
\usepackage{graphicx}
\usepackage{cite}
\usepackage{physics}
\usepackage{amssymb}
\usepackage{tabularx,booktabs}
\usepackage{mathtools}
\usepackage{multirow}
\usepackage{threeparttable}
\usepackage{MnSymbol}
\ifCLASSINFOpdf
\else
\fi

\usepackage[square, numbers, comma, sort & compress]{natbib} 
\usepackage{amsmath}
\usepackage[super]{nth}
\usepackage[justification=centering]{caption}
\usepackage{scalerel,stackengine}
\usepackage{xcolor}

\usepackage[printwatermark]{xwatermark}
\begin{document}

\title{Equivalence of space and time-bins  in DPS-QKD}

\author{\IEEEauthorblockN{Gautam  Shaw\IEEEauthorrefmark{1},
Shyam Sridharan\IEEEauthorrefmark{1}, 
Shashank Ranu\IEEEauthorrefmark{2}\IEEEauthorrefmark{1},
Foram Shingala\IEEEauthorrefmark{1}, 
Prabha Mandayam\IEEEauthorrefmark{2}, and
Anil Prabhakar\IEEEauthorrefmark{1} }

\IEEEauthorblockA{\IEEEauthorrefmark{1}Department of Electrical Engineering, IIT Madras, Chennai, India}\\
\IEEEauthorblockA{\IEEEauthorrefmark{2}Department of Physics, IIT Madras, Chennai, India}
\thanks{Gautam  Shaw is with the Department
of Electrical Engineering, Indian Institute of Technology, Madras,
e-mail: ee15d047@ee.iitm.ac.in.}% <-this % stops a space
\thanks{}% <-this % stops a space
\thanks{}}

\markboth{}%
{IEEE Journals}





\maketitle

\begin{abstract}

Key generation efficiency, and security, in DPS-QKD improve with an increase in the number of path delays or time-bin superpositions. We demonstrate the implementation of superposition states using time-bins, and establish an equivalence with  path-based superposition, thus yielding a simpler implementation of higher-order superposition states for differential phase-shift quantum key distribution (DPS-QKD). We set up  DPS-QKD, over 105~km of single mode optical fiber, with a quantum bit error rate of less than 15$\%$ at a secure key rate of 2~kbps. With temporal guard bands, the QBER reduced to less than 10$\%$, but with a 20$\%$ reduction in the key rate. 
\end{abstract}


\IEEEpeerreviewmaketitle

\section{Introduction}
%\IEEEPARstart{Q}
Quantum key distribution (QKD) enables secure key exchange between authenticated users, Alice and Bob, by relying on two aspects of quantum mechanics, Heisenberg's uncertainty principle and the no-cloning theorem~\cite{gisin2002quantum}. When an adversary, Eve, attempts to steal information from the quantum channel, she also inevitably introduces disturbances in the channel and reveals herself. Since the first   proposal by Bennett and Brassard in 1984 \cite{bennett1984proceedings}, there have been a variety of QKD protocols, both proposed and implemented \cite{liu2010decoy, stucki2007coherent,ekert1991quantum,inoue2002differential}. Long distance field demonstrations of QKD  mostly use discrete variables, some with active stabilization to mitigate environmental fluctuations~\cite{wang2014field}. In Appendix \ref{appendix}, we provide the reader with a quick summary of key rates and channel lengths for a few recent implementations of QKD. 

This article aims to establish the equivalence between a spatial and temporal generation of a superposition state for use in a differential phase-shift quantum key distribution (DPS-QKD) system. 
DPS-QKD as proposed by Inoue  et  al., is simple to implement and robust against slowly varying environmental fluctuations~\cite{inoue2002differential,inoue2005robustness}. DPS-QKD uses a pair of phases $\Phi$ = \{0, $\pi$\} to generate non-orthogonal states that cannot be distinguished with absolute certainty using a single measurement~\cite{bennett1992experimental}. A theoretical security proof of the DPS protocol, with single photons and weak coherent sources, was established under the assumption that Eve is restricted to individual attacks and also it was  concluded that individual attacks are more powerful than sequential attacks~\cite{waks2006security, wen2009unconditional}. An efficient phase encoding quantum key generation scheme, with narrow band heralded photons, was proposed by Yan et al., where key information is carried by the phase modulation directly on the single-photon temporal waveform \cite{hui2011efficient,liu2013differential}. Time-bin qubits, composed of temporal modes with weak coherent sources, are  effective constituents  to  build a robust and simple  QKD system with a  high secure key rate \cite{boaron2018simple}. 
To increase the secure key rate, with minimum resources, researchers have used two-dimensional and  four-dimensional QKD protocols with time-bin and phase encoding~\cite{vagniluca2020efficient}.
Dellantonio  et al.  proposed two equivalent high dimensional MDI-QKD methods: space and time-bin encoding, which uses space to encode information in different paths and time-slots to encode qudits \cite{dellantonio2018high}.  In high dimensional QKD protocols, multiple bits of information are encoded on a single photon, hence, it increases the channel capacity  and is more robust against  channel noise.

The recently introduced round-robin differential phase-shift quantum key distribution (RR-DPS-QKD) scheme addresses the effects of environmental disturbances, and gives us an upper bound on our tolerance to error rates  with a bit error rate as high as $29\%$ \cite{guan2015experimental}. %In principle, RR-DPS QKD protocol can tolerate a bit error rate (BER) as high as $50\%$~\cite{guan2015experimental}. 
But such schemes requires the addition of  optical switches and delays that make Bob's set-up more complex~\cite{QU201943}.

In Sec.~\ref{Eresults}, we show that the two schemes, of spatial and temporal superposition, yield comparable key rates in kbps, with a \text{QBER} $<$ 0.2. However, time-bins are defined electronically, and are significantly easier to generate. The scheme does require more precise timing synchronization, and we have developed the means to characterize the photon arrival time at our detector to within 50 ps. We can implement DPS-QKD with a temporal multiplexed single photon detector. We further demonstrate that temporal filtering can reduce the QBER of our DPS-QKD implementation, but at a reduced secure key rate.

\section{Experimental Setup}
In the first  DPS-QKD proposal, a single photon was allowed to pass through a beam splitter, travel through different path delays and then  recombined to create a superposition state of the photon \cite{inoue2002differential}. However, this  scheme would encounter beam splitter losses and reduces the secure key rate. The more common version of DPS-QKD is one that uses weak coherent pulses (WCPs) \cite{inoue2003differential}.  With a path superposition of $N$-paths, and a relative phase of $\{0, \pi\}$ between paths, the photon can be in a superposition of $2^{N-1}$ states. Similar states can also be achieved with  time-bin superposition by adding a relative phase at $N-1$ locations within a single pulse. In this article, we define $2^{N-1}$ as $M$, and hence, refer to the 3-pulse DPS-QKD as a 4-state system. This notation will allow us to describe the creation of superposition states by temporal phase modulation.

\begin{figure}[h!]	
 	\centering

 	\includegraphics[width=8.4cm,height=5.6cm]{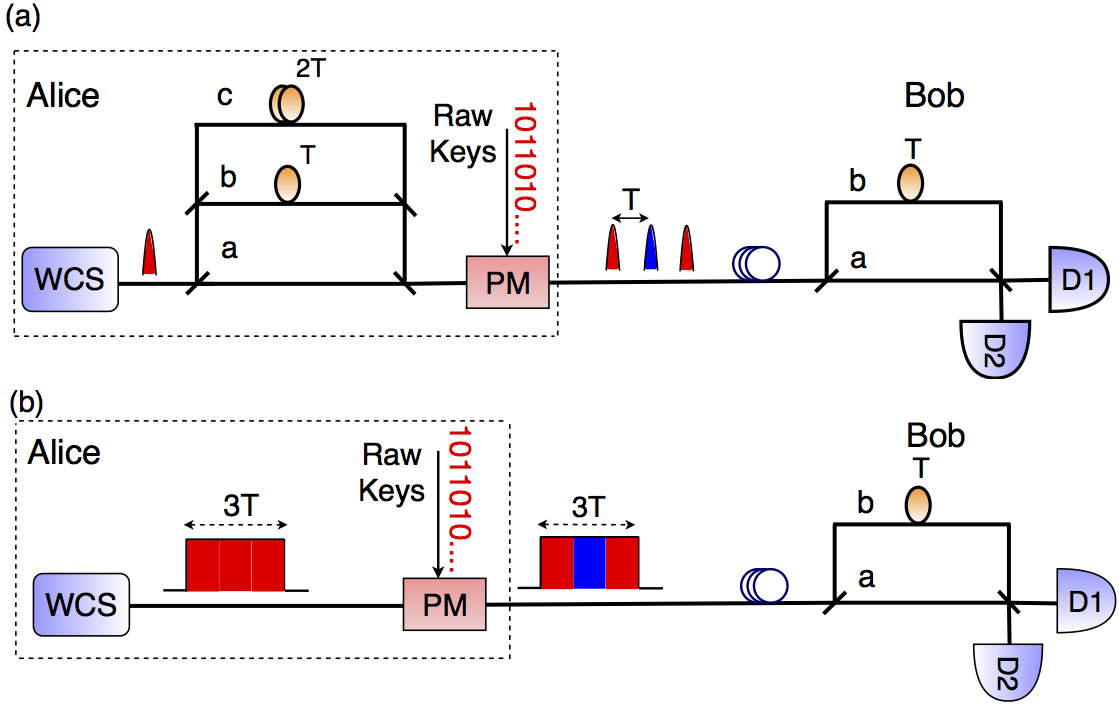}
 	\caption{4-state DPS-QKD schemes using (a) spatial and (b) time-bin superposition with a weak coherent source. WCS: Weak coherent source, PM: phase modulator, T: time, D1 and D2: single photon detectors.}
 	\label{fig:4 state DPS-QKD source}
 \end{figure}

We describe our experiments with 4-state DPS-QKD, using space and time-bin superposition of a weak coherent source, as shown in Fig.~\ref{fig:4 state DPS-QKD source}. The time-bin superposition scheme is easier to implement and control, and can be extended to an $M$-state DPS-QKD scheme without any additional hardware complexity.  When we use a superposition of 4 states, an intercept and resend (IR) attack by Eve introduces a $33\%$ error on the sifted key. We had previously reported that the 4-state DPS scheme is more secure against both IR and beam splitter attacks \cite{ranu2017security}. This percentage error increases to $50\%$ when 4-state DPS is extended to $M$-state DPS, but with ideal detectors ~\cite{ranu2017security}.
The methodology used to implement a weak coherent source (WCS) is described along with the rest of our experimental setup in Sec.~\ref{Exper_final_setup}. 

 \section{Key generation in DPS-QKD}
\begin{figure}[hbt!]
 \centering
 \centering
 \centerline{\includegraphics[width=1.0\columnwidth]{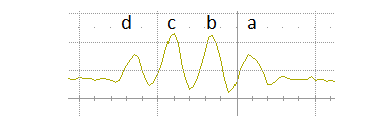}}
 \caption{Photodetector output after Alice's path superposition and Bob's DLI, captured with a diode laser source. The key is generated by the interference in time-bins $b$ and $c$.}
 \label{fg:4pulseDSO}
 \end{figure}

In our 4-state DPS-QKD implementation. Alice sends a single photon in a superposition of $3$ spatial paths to Bob. The probability of a photon traveling through one of the $3$ paths in  Alice's set-up is $1/3$.  The superposition state generated from Alice is represented as:
 \begin{align}
     \ket{\Psi} &= \frac{1}{\sqrt{3}}\left[{\ket{1}_{\text{a}}\ket{0}_{\text{b}}\ket{0}_{\text{c}}\pm  \ket{0}_{\text{a}}\ket{1}_{\text{b}}\ket{0}_{\text{c}}\pm \ket{0}_{\text{a}}\ket{0}_{\text{b}}\ket{1}_{\text{c}}}\right] \\
     & \triangleq \frac{1}{\sqrt{3}}\left[\ket{100}_{\text{abc}} \pm  \ket{010}_{\text{abc}} \pm \ket{001}_{\text{abc}}\right]
     \label{eq:psi}
 \end{align}
 where the paths a, b, c also represent time-bins. $\ket{\Psi}$ is passed through a delay line interferometer (DLI) at Bob's site.  As a result, the photon is now in a superposition of $4$ time-bins. The first and last  time-bins do not contain encoded phase difference information, whereas  the  $2$ central time-bins contribute to the key generation. The 4 time-bins can also be observed classically, but at higher photon numbers, as shown in Fig.~\ref{fg:4pulseDSO}. Alice now encodes her random key bit  as a random phase $\phi=\{0,\pi\}$ between successive time-bins. Bob extracts the key information using a DLI and two single-photon detectors. Eve's intercept and resend attack introduces an error of 33$\%$ in the sifted key in the 4-state DPS compared to the 25$\%$ error when using a train of WCPs in conventional DPS-QKD.  
 
\begin{figure}[hbt!]	
 	\centering
 	\includegraphics[width=8.8cm, height =5cm]{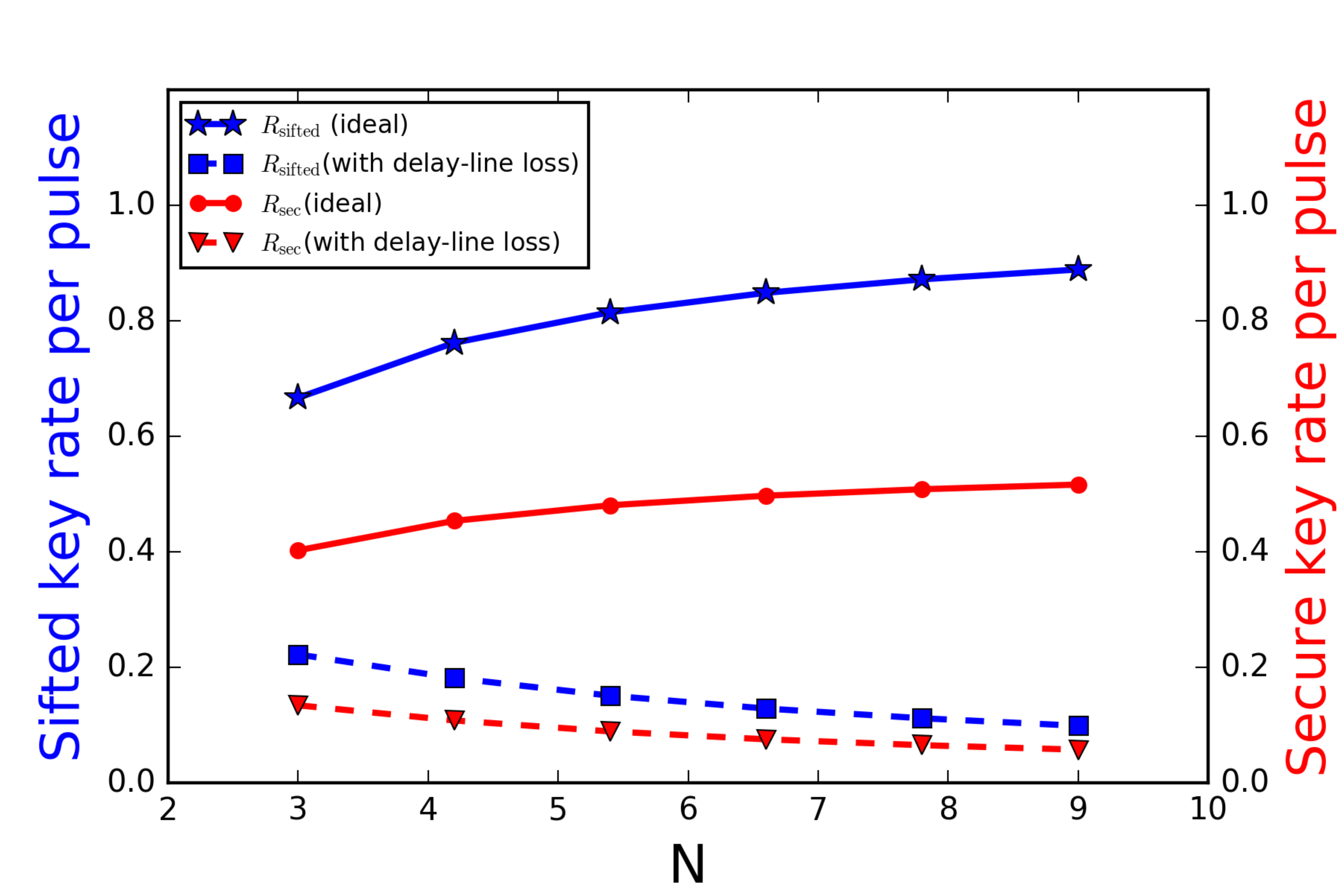}
 	\caption{Estimates for the sifted and secure key rate for $M$-state DPS-QKD, with ideal detectors.}
 	\label{fig:sifted key rate}
\end{figure}

  %A photon is detected in the first (last) time slot if it travels through first (third) path in  Alice's 3 pulse set-up and shorter (longer) arm in Bob's DLI. The probability of detection of a photon in the  $\nth{1}$ and $\nth{4}$ time slot is $\frac{1}{6}$+$\frac{1}{6}=\frac{1}{3}$.  Thus, the sifted key rate $(R_\text{sifted})$ is equal to $\frac{2}{3}$. Similarly, if Alice transmits $n$ photons where each photon is in a superposition of $N$ time-bins, only $n\left[\frac{N-1}{N}\right]$ photons contribute to the final keys, and hence the sifted key rate $(R_\text{Sifted})$ is equal to $\frac{N-1}{N}$~\cite{ranu2017security}. For a higher value of $N$, the sifted key rate converges towards 1 as shown in Fig.~\ref{fig:sifted key rate}. We also show the dependence of secure key rate on the number of delays/time-bins in Alice's setup. The secure key rate is given as~\cite{diamanti2006100}
The secure key rate ($R_{\text{sec}}$) is estimated from sifted key rate ($R_{\text{sifted}}$) as \begin{equation}
     R_{\text{sec}}=R_{\text{sifted}}\left[\tau-f(e)h(e)\right],
\end{equation}
where $\tau$ is the shrinking factor, $f(e)$ captures the inefficiency of the error correcting code, and $h(e)$ is the binary Shannon entropy. The error rate, $e$,  depends upon dark counts and other system imperfections and $\tau$ captures Eve's knowledge of the key. 
If we assume Eve's attack to be limited to the IR and beamsplitter attacks, increasing $N$ changes the efficacy of the attacks, thus making $\tau$ a function of $N$~\cite{ranu2017security}. Hence, a secure key rate that depends on both $R_{\text{sifted}}$ and $\tau$, varies with $N$ as shown in Fig.~\ref{fig:sifted key rate}.
 
Experimentally, the generation of a superposition state can be realized spatially using passive beam splitters (or beam combiners). However, passive beam splitters have insertion losses and the sifted key rate is reduced by a factor of $N$, thus making the implementation inefficient. An implementation with $N=3$ will create $M=4$  non-orthogonal states that Alice uses to transmit the key. Keeping in mind the ease of implementing time-bin superposition, we advocate temporal bins using phase modulation, over different spatial paths. This would potentially allow us to use $N>3$ and obtain a higher sifted key rate, as seen in Fig.~\ref{fig:sifted key rate}. However, with non-ideal SPDs, the optimal value against all attacks is $N =3$ \cite{wen2009unconditional}. We observe this by the increase in QBER due to afterpulsing, within the same gate pulse, as $N$ increases \cite{SHAW2022168280}.

 \section{Experimental Results}
 \label{Eresults}
The  state at the output port of Bob's DLI, consists of a single photon in one of 4 time-bins. A correct identification of these time-bins needs an accurate temporal characterization of the single photon detector (SPD), and its associated electronics. 

\subsection{Experimental set-up}
\label{Exper_final_setup}
Alice's set-up consists of  a continuous laser source  at 1550.12 nm and a RF pulse generator. A train of electrical pulses, having a pulse width of 500~ps and a time period of 32~ns, is applied to a 10~GHz intensity modulator (IM). 
The bias voltage to the IM was optimized  to get a modulation extinction ratio of  more than 14~dB. The resultant optical pulses  were then attenuated, using  a variable optical attenuator (VOA), to a mean photon number  $\mu \sim 0.1$, and the photons were  sent directly to a gated SPD. We observed 31~K~counts/s when the gate window was synchronized with the photon arrival time. This reduced to around 1~K~counts/s  when the gate was out of sync with the arrival time of the photons.

\begin{figure}[hbt!]
	\centering
	\includegraphics[width=9cm, height =5.2cm]{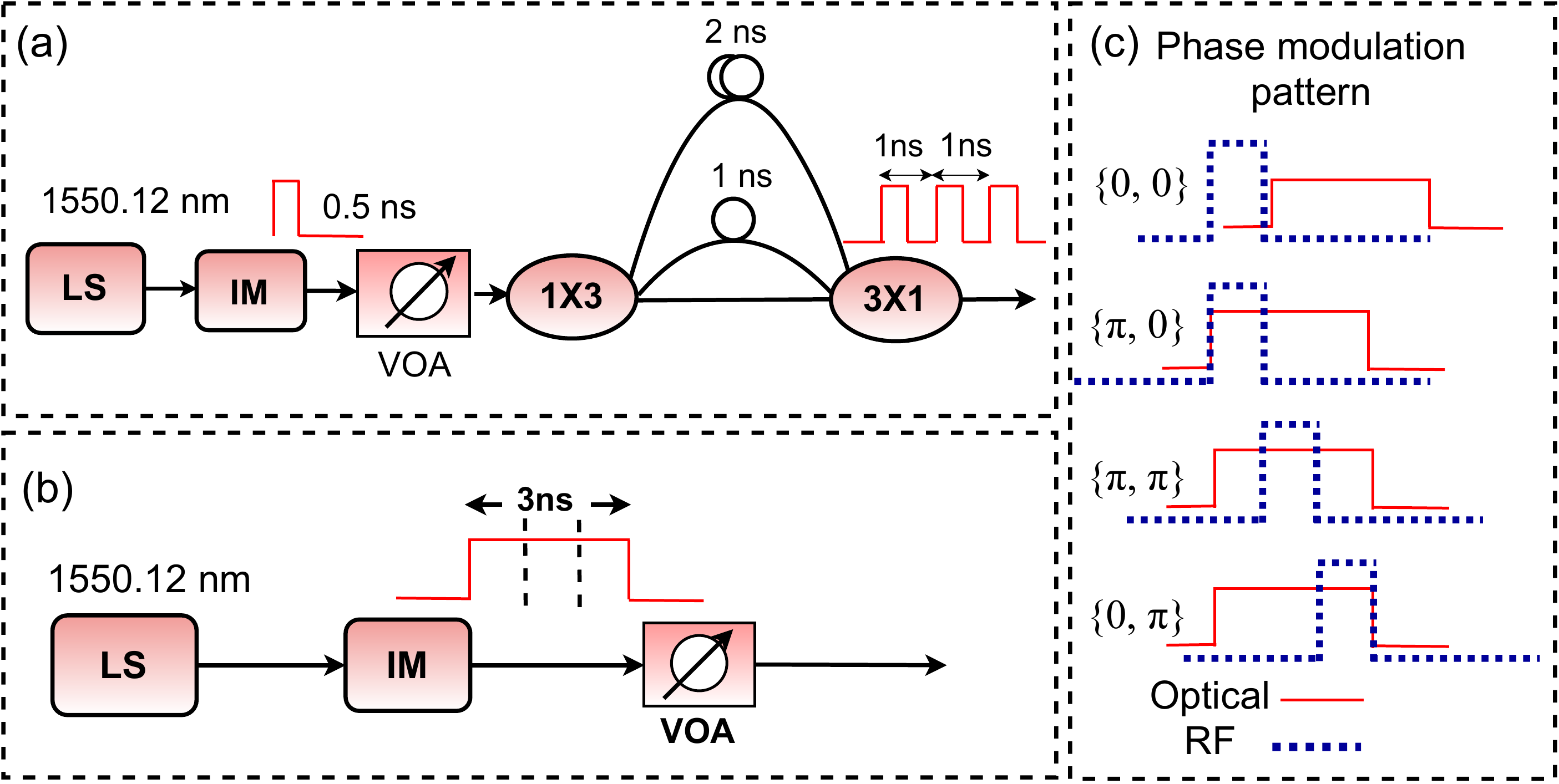}
	\centering
	\caption{Weak coherent sources for (a) spatial, and (b) time-bin superposition. LS: laser source, IM: intensity modulator, VOA: variable optical attenuator. (c) Phase modulation pattern}
    \label{two different configurations}
\end{figure}

\begin{figure}[hbt!]	
	\centering
	\includegraphics[width=9.0cm, height =5.9cm]{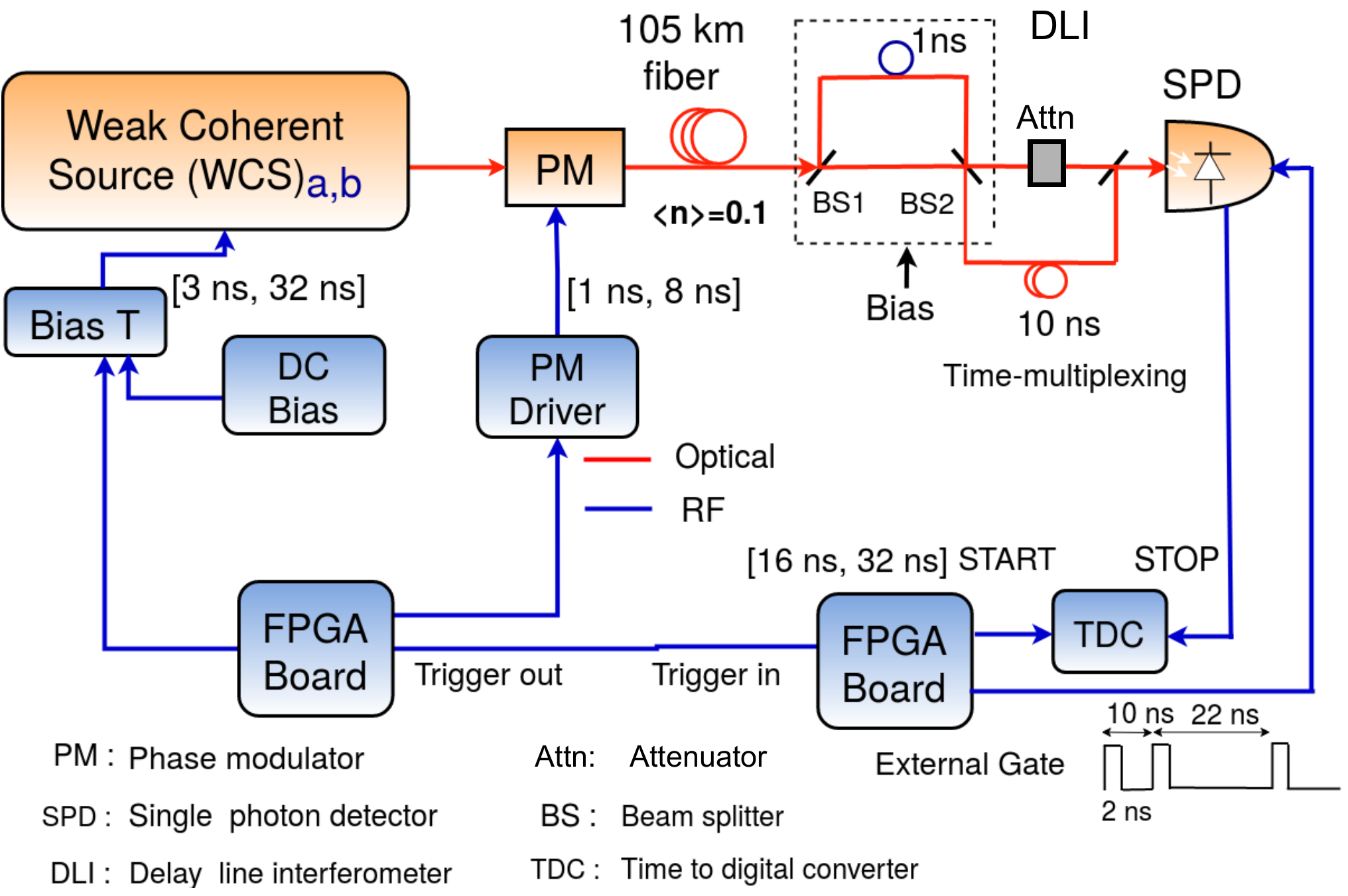}
	\caption{4-state DPS-QKD experimental set-up. Red and blue coloured blocks represent electro-optic and electrical (RF)  devices used in the test-bed respectively. A detailed schematic of the weak coherent source is shown in Fig. \ref{two different configurations}}
   \label{fig:dps-qkd expt set-up}
\end{figure}

Two different source configurations for path and time-bin superposition, as shown in Fig.~\ref{two different configurations}, were then used in the DPS-QKD experimental set-up shown in Fig.~\ref{fig:dps-qkd expt set-up}. 
In Fig.~\ref{two different configurations}(a), weak coherent pulses are passed through 1$\times$3 and 3$\times$1 beam splitter-coupler combination  so that photons coming out from $3 \times 1$ coupler are in superposition of three paths before being passed through the phase modulator (PM),  shown in Fig.~\ref{fig:dps-qkd expt set-up}. In Fig.~\ref{two different configurations}(b), a 3~ns pulse coming out of the intensity modulator (IM) is attenuated and acts as a source. Phase modulation can be introduced on this pulse, as shown in Fig. \ref{two different configurations}(c).

One problem with using two independent detectors to differentiate between 0 and 1 bits is that the detectors are not identical, and will typically have different quantum efficiencies. We mitigate this by using time-multiplexing and capture photon arrival  times from both output ports of the DLI.  A fiber delay of 10~ns was added at one of the output ports of the DLI. To equalize the loss in both paths from output ports of DLI, an attenuator of 0.5 dB was added in short path. Both ports were then multiplexed using a $2 \times 1$  coupler and sent to a SPD. This technique provides a cost-effective configuration since one SPD is enough to extract timing instant information. Unfortunately, half of the photons are lost due to the $2 \times 1$ combiner before the SPD. But, since we can generate WCPs at GHz rates, we are limited only by the hold-off time on the SPD and do not perceive any disadvantage to using a time-multiplexed configuration with a single SPD. Instead, using a single detector has the advantage of providing an equal sensitivity on both constructive and destructive interference ports of the DLI. An IM with a high extinction ratio reduces false detections during the 10 ns off time in the time-multiplexed detection scheme. 

A field programmable gate array (FPGA) is triggered synchronous to the pulse generator and it is configured to generate control signals for the SPD, TDC  and a modulating signal (RF pulses) for the PM. Phase encoding patterns \{0, 0\}, \{$\pi$, 0\}, \{0, $\pi$\} and \{$\pi$, $\pi$\} are realized by applying  RF pulses to the phase modulator  synchronous to the three different time locations  within a 3 ns temporal wave packet, as shown in Fig. \ref{two different configurations}(c).  The FPGA also provides a variable gate delay  to synchronize the full systems, and to identify the interference slots. We recorded the photon arrival times  by varying the RF delay to the PM for a fixed gate delay.  Sifted key generation and QBER measurements for both space and time multiplexed schemes were obtained after integrating a TDC and a time-stamp module in the FPGA.

\section{Results and discussion}

A sifted key was derived after counting the TDC output, combined with that from a time-stamp module. 
\subsection{Key generation}
\label{Key generation}
\begin{figure}[hbt!]	
	\centering
	\includegraphics[width=8.0cm, height =8.7cm]{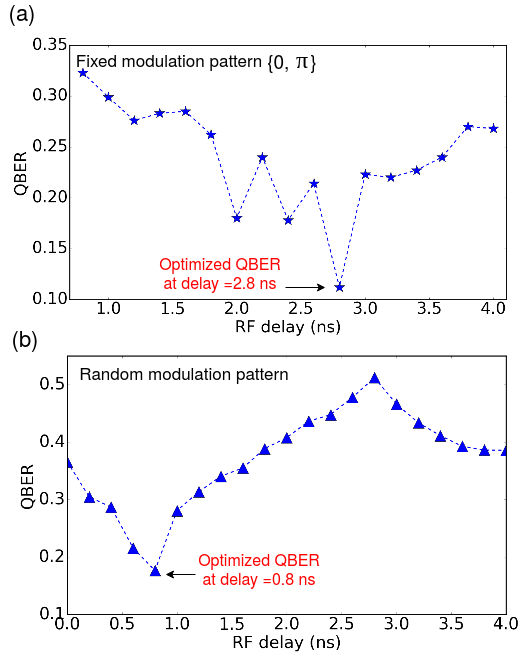}
	\caption{Optimization of QBER by adjusting the timing of the applied phase (a) with fixed phase modulation pattern (b) with random phase modulation pattern}
    \label{optimized QBER}
\end{figure}

A final key rate  of 21~kbps and 10~kbps was achieved in the time-bin superposition and path superposition schemes, with a QBER of 0.17  and 0.21,  respectively, over 30~km of optical fiber. By further optimizing the DLI bias voltage and the control parameters of the SPD, we were able to observe a QBER of 0.12, shown pictorially in Fig.~\ref{optimized QBER}(a), in the time-bin scheme. 

\begin{figure}[!htb]	
	\centering
	\includegraphics[width=9cm, height =9.5cm]{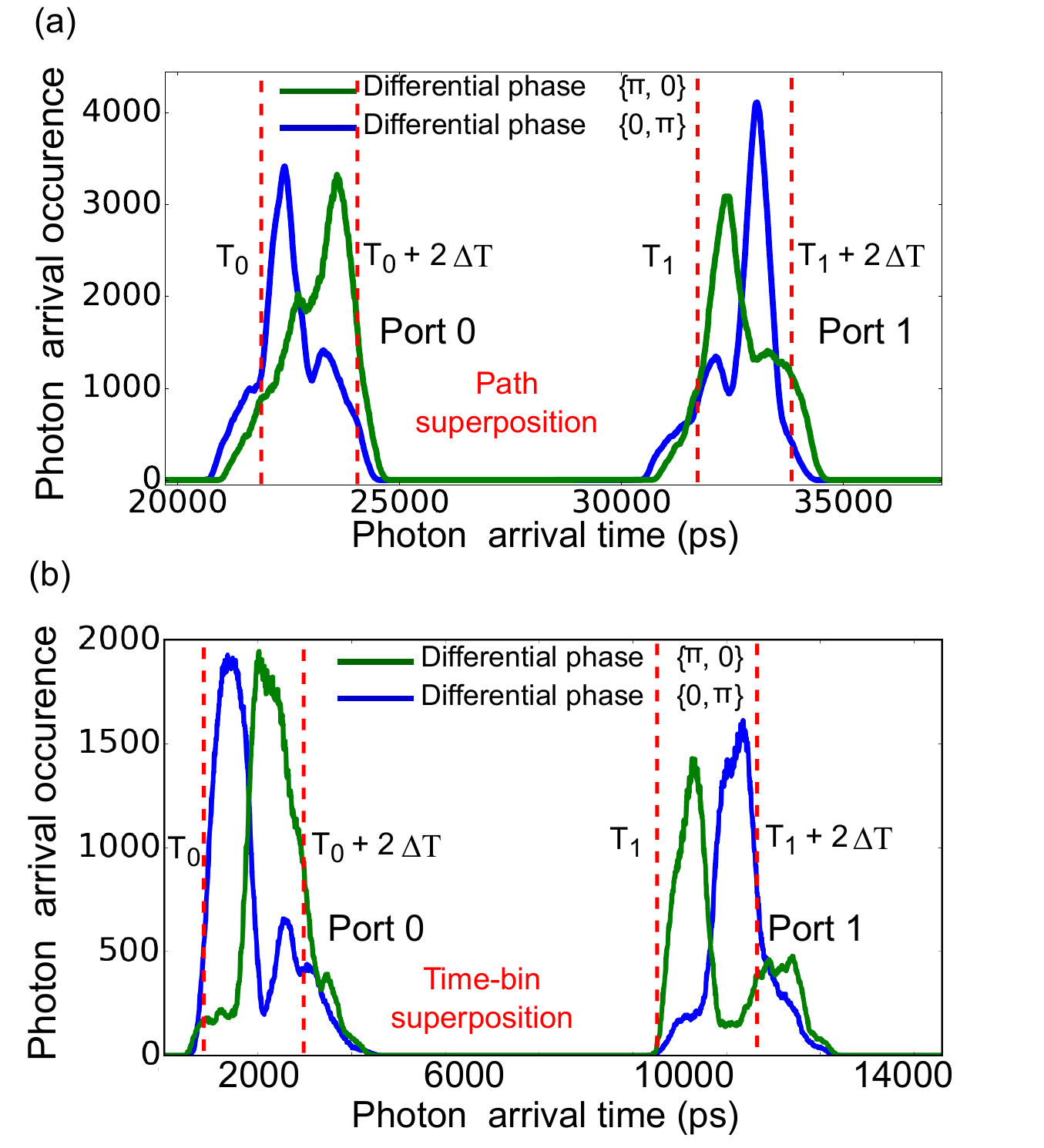}
	\centering
	\caption{Photon arrival time distribution  for (a) path superposition  and  (b) time-bin superposition}
    \label{Photon arrival time distribution in 4-state DPS-QKD}
\end{figure}

With reference to Figs. \ref{Photon arrival time distribution in 4-state DPS-QKD} (a) and (b), the QBER is defined as,
\begin{equation}
\text{QBER} = \frac{C_{01}+C_{10}}{C_{00}+C_{10}+C_{01}+C_{11}}
\end{equation}
where
%where $C_{01}$($C_{10}$) represents the error counts at constructive port (destructive port) of DLI for $j^{th}$ time slot, when Alice's transmitted raw key is '1'('0') at $j^{th}$ time slot.  $C_{01}$ and $C_{10}$ are calculated as follows:
\begin{align}
C_{\text{pq}}&=\sum_{t=T_{\text{start}}}^{T_{\text{stop}}} c_{t}
\end{align}
%where $\Delta T$ is the width of the time slot, $c_{t}$ represents the photon arrival occurrence at a time $t$ and $T_1$-$T_0$ = 10 ns. 
%Table for calculating QBER using histogram
represents the counts at the $p^{\text{th}}$ port of the DLI, when Alice's transmitted raw key is $q$. $T_{\text{start}}$ and $T_{\text{stop}}$ are determined for each case from Table. \ref{table:QBER calculation}.

\begin{table}[tbh]
\caption{Classification of photon counts} 
\centering % used for centering table
\begin{tabular}{|l|l|l|l|}
%\hline
%\multicolumn{4}{|c|}{Team sheet} \\
\hline
Phase pattern & $C_{\text{pq}}$ & $T_{\text{start}}$ & $T_{\text{stop}}$ \\ \hline
\multirow{4}{*}{\{0, 0\}} & $C_{00}$ & $T_{0}$  & $T_{0}+2\Delta T$  \\
 & $C_{01}$ & $--$  & $--$   \\
 & $C_{10}$ & $T_{1}$  & $T_{1}+2\Delta T$  \\
 & $C_{11}$ & $--$  & $--$ \\ \hline
\multirow{4}{*}{\{0, $\pi$\}} & $C_{00}$ & $T_{0}$  & $T_{0}+\Delta T$  \\
 & $C_{01}$ & $T_{0}+\Delta T$  & $T_{0}+2\Delta T$  \\
 & $C_{10}$ & $T_{1}$  & $T_{1}+\Delta T$  \\
 & $C_{11}$ & $T_{1}+\Delta T$  & $T_{1}+2\Delta T$  \\ \hline
\multirow{4}{*}{\{$\pi$, 0\}} & $C_{00}$ & $T_{0}+\Delta T$  & $T_{0}+2\Delta T$ \\
 & $C_{01}$ & $T_{0}$  & $T_{0}+\Delta T$  \\
 & $C_{10}$ & $T_{1}+\Delta T$  & $T_{1}+2\Delta T$  \\
 & $C_{11}$ & $T_{1}$  & $T_{1}+\Delta T$ \\ \hline
\multirow{4}{*}{\{$\pi$, $\pi$\}} & $C_{00}$ & $--$  & $--$  \\
 & $C_{01}$ & $T_{0}$  & $T_{0}+2\Delta T$  \\
 & $C_{10}$ & $--$  & $--$  \\
 & $C_{11}$ & $T_{1}$  & $T_{1}+2\Delta T$  \\ \hline
%\hline
\end{tabular}
\label{table:QBER calculation} % is used to refer this table in the text
\end{table}

We recovered the sifted key and extracted the QBER by directly comparing the sender's keys with the receiver's. This approach was used to optimize the RF delay and appropriately insert a phase shift every 1~ns within the 3~ns optical pulse, using a fixed pattern of \{0, $\pi$\}. Although the phase pattern was fixed, with a low mean photon number, channel loss, and a detector efficiency $\eta\sim0.1$, we only detect a random bit pattern after the delay line interferometer.

The QKD testbed was also used to investigate the effect of excess bias voltage, gate width and hold-off time on the dark count rate (DCR) and afterpulse noises of a gated InGaAs single-photon detector (SPD) \cite{SHAW2022168280}. This helped in improvising the QBER for all 4 possible phase modulation patterns. We achieved a QBER of 0.12  for a fixed phase pattern \{0, $\pi$\}, as shown in Fig. \ref{optimized QBER} (a). Other patterns \{0, 0\}, \{$\pi$, 0\} and  \{$\pi$, $\pi$\}  yielded a QBER of 0.9, 0.14 and 0.27 respectively, while  a random pattern yielded a minimum QBER of 0.17 as shown in Fig. \ref{optimized QBER}(b).  

\begin{figure}[hbt!]	
	\centering
	\includegraphics[width=8.2cm, height =10.5cm]{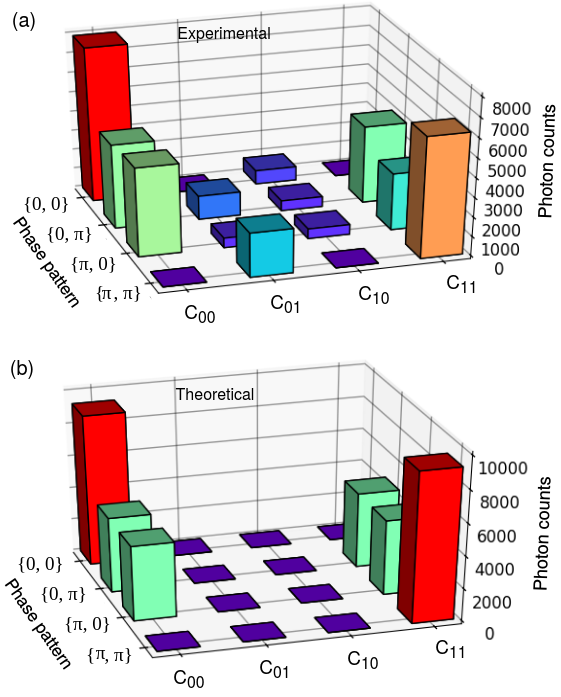}
	\caption{Time-bin based differential-phase decoding by collecting photon arrival time for all four phase modulation states (a) experimental measurement (b) theoretical estimation}
    \label{fg:tomograph_and_states}
\end{figure}

The sifted key generation rate in our time-bin superposition DPS system can be written as~\cite{diamanti2006100}
\begin{equation}
   R_{\text{sifted}}= r_{\text{p}} \mu \eta T_L e^{(-r_{\text{p}} \mu \eta T_L \tau_{\text{H}})} 
   \label{sifted equation}
\end{equation}
where the variables used are defined in Table \ref{table:variables_equation}. 

\begin{table}[tbh]
\caption{Variables used in sifted key rate (\ref{sifted equation})} % title of Table
\centering % used for centering table
\fontsize{9}{11}\selectfont
\begin{tabular}{|p{1.2cm} p{5.8cm}|} % centered columns (4 columns)
\hline %inserts double horizontal lines
Variables & Description  \\
\hline % inserts single horizontal line
\hline
$r_{\text{p}}$  & Pulse repetition rate \\
$\alpha$ & Attenuation constant of a single mode fiber \\
$\mu$ & Mean photon number per pulse \\
$T_L$ & $10^{-(\frac{\alpha L+ I_L}{10})}$ (overall transmission efficiency of quantum channel)\\
$\eta$ & Detection efficiency of SPD \\
$T_{\text{hold}}$ & Hold-off time of SPD \\
% [1ex] adds vertical space
\hline %inserts single line
\end{tabular}
\label{table:variables_equation} % is used to refer this table in the text
\end{table}

$T_L$ consists of fiber loss due to attenuation ($\alpha \approx 0.2$ dB/km) and the net insertion loss ($I_L$) of the DLI and coupler. Referring to (\ref{sifted equation}), the values of $r_{\text{p}}$, $\eta$ and $\tau_{\text{H}}$ are 62.5~Mbps, 10$\%$ and 10 $\mu$s respectively. The exponential term in (\ref{sifted equation}) approaches 1 for a transmitted pulse rate of 62.5~Mbps, with  a hold-off time of  $10\,\mu$s, and $R_\text{sifted}$ decreases linearly with distance. However, the exponent becomes significant for higher transmitted pulse rates, typically  $r_{\text{p}} > 1\,\text{Gbps}$. 

\begin{figure}[hbt!]	
	\centering
	\includegraphics[width=8.8cm, height =5.2cm]{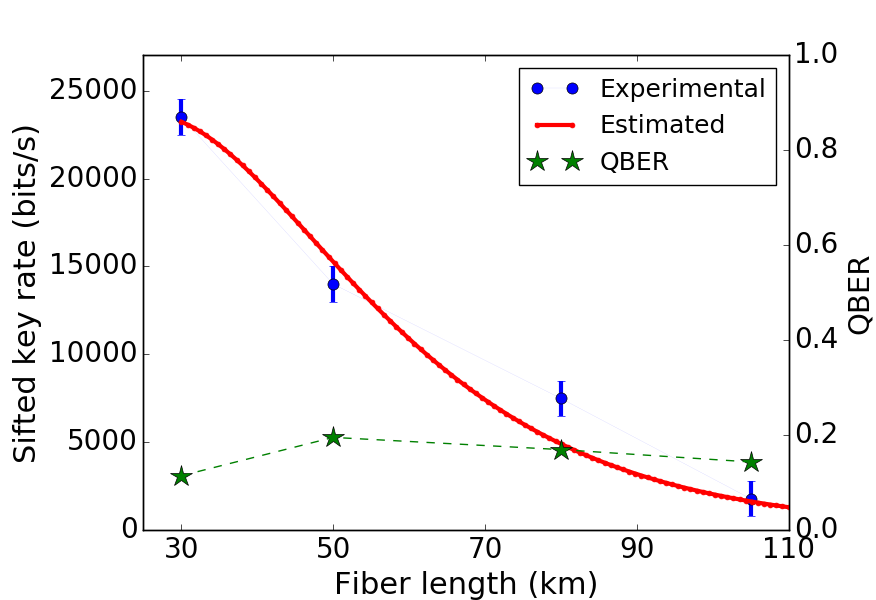}
	\caption{Sifted key rate (estimated and experimental) and measured QBER  as a function of channel length}
    \label{fg:qber_keyrate}
\end{figure}
As we observe in Fig.~\ref{fg:qber_keyrate}, the experimental data fits well to (\ref{sifted equation}), and we estimate $\mu \approx 0.17$.  At a fiber length of 30~km, we achieved a sifted key generation rate of 21~kbps with a QBER of 11.5 $\%$. We then extended our experiment to  105~km of fiber, and observed the sifted key rate drop to about 2~kbps with a QBER of 14.4 $\%$, as shown in Fig.~\ref{fg:qber_keyrate}.

\begin{figure}[hbt!]	
	\centering
	\includegraphics[width=8.8cm, height =5.2cm]{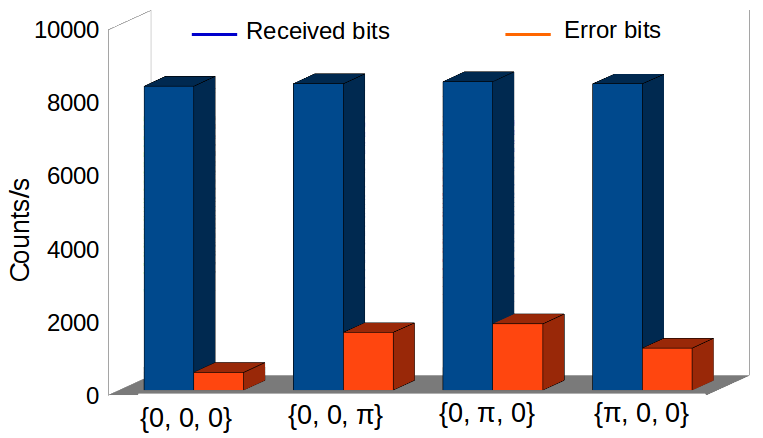}
	\caption{Differential phase decoding in 8-state DPS-QKD. QBER for transmitted phase modulation states :\{0, 0, 0\}, \{0, 0, $\pi$\}, \{0, $\pi$, 0\} and \{$\pi$, 0, 0\} are 0.06, 0.20, 0.22 and 0.14 respectively}
    \label{fg:qber_8 state}
\end{figure}
We further extended our 4-state DPS experiments to realize 8-state DPS,  with a DLI having free spectral range of 2.5 GHz. The phase modulation transition rate was enhanced from 1 GHz to 2.5 GHz within a photon wave packet of width 1.6 ns, where each time-bin size is 0.4 ns. Based on the photon arrival time within 1.2 ns, the QBER was calculated for various phase modulation states, as shown in Fig. \ref{fg:qber_8 state}. 

\subsection{Temporal filtering of time-stamps}
\begin{figure}[hbt!]	
	\centering
	\includegraphics[width=8.7cm, height =11.0cm]{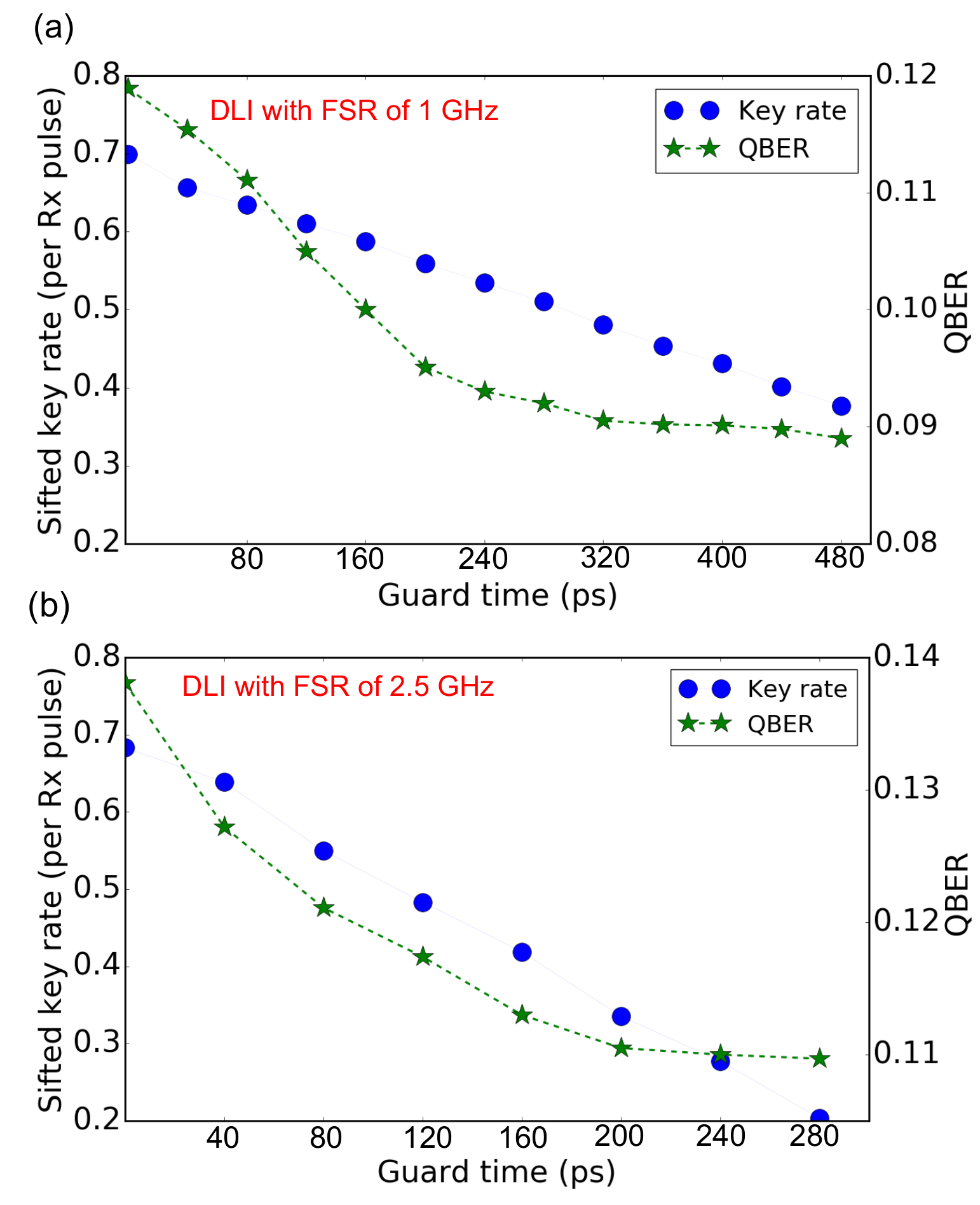}
	\caption{Effect of temporal filtering on sifted key rate and QBER, (a) with DLI of FSR 1 GHz (b) with DLI of FSR 2.5 GHz.  At higher guard time, QBER reduces at a cost of reduced sifted key rate per $R_{\text{x}}$ detection.}
    \label{fg:qber_guardband}
\end{figure}

After identification of the optimized gate and RF delay, we introduce a guard window between time-bins at the output state of Bob's set-up \cite{gautam_spcom_2022}. In temporal filtering, while we discard the time-stamps collected at the selected guard time, we lose a fraction of bits in sifted keys. However, this method reduces the QBER of a system \cite{kupko2020tools}, but at a cost of a reduced sifted key rate, as shown in Fig. \ref{fg:qber_guardband} (a) and (b). 
Table~\ref{table:error_list} lists the  source of errors, contributing in QBER for our QKD test-bed. Error due to DCR, afterpulse probability and timing jitter were estimated from our previous work on gated SPD characterization \cite{shaw20193}. 
\label{Performance optimization}
\begin{table}[tbh]
\caption{Potential sources of errors in QKD test-bed} 
\begin{threeparttable}[t]
\centering % used for centering table
%\fontsize{9}{11}\selectfont
\begin{tabular}{|l|l|l|l|}
%{p{2.8cm} p{1.0cm} p{3.3cm}} % centered columns (4 columns)
\hline %inserts double horizontal lines
Description & Source  & \multicolumn{2}{c|}{Error contribution ($\%$)}  \\
  &  & $\Delta T$= 1 ns & $\Delta T$= 0.4 ns \\
\hline
\hline % inserts single horizontal line
Dark count rate (DCR)& SPD & 0.33 & 0.33 \\
Afterpulse effect & SPD  & 1.5 & 1.5\\
Extinction ratio  & IM & 1.6 & 1.6\\
Timing jitter & SPD  & 5.0 & 12.5 \\
Imperfect visibility & DLI & 4.0 & 2.0\\
Rise/fall time of PM pattern$^{\star}$  & PM & 2.1 & 5.25\\
\hline
\end{tabular}
\begin{tablenotes}
     \item[$^{\star}$] From data sheet.
   \end{tablenotes}
\end{threeparttable}
\label{table:error_list} % is used to refer this table in the text
\end{table}
The interferometric visibility ($V$) of two DLI with FSR of 1 GHz and 2.5 GHz are 92$\%$ and 96$\%$ respectively. The QBER introduced into the system due to $V$  is  $(1-V)$/$2$. Referring  to Table~\ref{table:error_list}, the major sources of QBER are timing jitter of SPD and phase modulation rise/fall time, which can be reduced by temporal filtering method.  
For a guard time of 200 ps, 20 $\%$ key bits are discarded, while the QBER is reduced from 0.12  to 0.09 in a  system with a DLI of FSR 1 GHz, as shown in Fig. \ref{fg:qber_guardband} (a). For a DLI with 2.5 GHz FSR, we  discard 20 $\%$ key bits, to reduce QBER from 0.14  to 0.12, as shown in Fig. \ref{fg:qber_guardband} (b). The reduction in QBER for a larger FSR is only marginal. This highlights the challenge of moving to a  DLI with a higher FSR.
After estimating and optimizing QBER, we followed error correction and privacy amplification to generate secure keys. However, our focus is on QBER measurement and sifted key generation, hence error correction \cite{andrew2015construction} and privacy amplification are not discussed in this paper.

\section{Conclusion and perspectives}
We have presented two different experimental approaches (a) spatial superposition and (b) time-bin superposition, to realize a  4-state DPS-QKD over a  105~km quantum channel.  We observe that the time-bin superposition scheme is more efficient and easier to implement and can be extended to an $M$-state DPS-QKD system. After optimization of various parameters, we achieved a sifted key rate of around 21~kbps with QBER of 0.12 over 30~km of fiber. We then extended our time-bin superposition based DPS-QKD system to 105~km of optical fibre, and achieved a sifted key rate of 2~kbps while maintaining a QBER of 0.14. We also applied temporal filtering method in our DPS-QKD system, and shown that it helps to reduce the QBER, back to 0.12, but at a cost of reduced sifted key rate.

\section*{Acknowledgment}
This work was supported by Ministry of Human Resources and Development (MHRD) vide sanction no. 35-8/2017-TS. 
\appendix
%\section{}
\label{appendix}
\begin{table}[htbp!]
\caption{Decoy state implementations \cite{boaron2018secure, yuan201810, frohlich2017long}} % title of Table
\begin{threeparttable}[t]
%\fontsize{9}{11}\selectfont
\centering % used for centering table
\begin{tabular}{|p{2.3cm}| p{1.2cm}| p{1.2cm}| p{0.9cm}| p{0.9cm}|} % centered columns (4 columns)
\hline %inserts double horizontal lines
Author, & Protocol & Encoding  & Channel & Key rate \\
Year    &         & scheme & length & bits/s \\
\hline % inserts single horizontal line
Frohlich et al., 2017 & BB84 & Phase & 240 km& 8.4 \\
Boaron et al., 2018 & Simplified BB84  & Time-bin & 421 km& 6.5\\
Yuan et al., 2018 & BB84 variant & Phase & 10 km& 13.7 M \\
 % [1ex] adds vertical space
\hline %inserts single line
\end{tabular}
\end{threeparttable}
\label{table:nonlin1} % is used to refer this table in the text
\end{table}

\begin{table}[htbp!]
\caption{Measurement-device-independent (MDI) QKD implementations \cite{yin2016measurement, tang2016experimental, comandar2016quantum, wang2017measurement, valivarthi2017cost, liu2019experimental, wei2020high, zhou2021reference}} % title of Table
\begin{threeparttable}[t]
\centering % used for centering table
%\fontsize{9}{11}\selectfont
\begin{tabular}{|p{2.5cm}| p{1.2cm}| p{1.3cm}| p{0.9cm}| p{0.8cm}|} % centered columns (4 columns)
\hline %inserts double horizontal lines

Author, & Protocol & Encoding  & Channel & Key rate \\
Year    &         & scheme & length & bits/s \\
\hline % inserts single horizontal line
Yin et al., 2016  & Decoy state MDI   & Time-bin & 404 km & 0.00032\\
Tang et al., 2016  & BB84 & Polarisation & 40 km& 10 \\
Comandar et al., 2016  & BB84 & Polarisation & 102 km& 4.6 K \\
Wang et al., 2016  &\text{RFI}\tnote{$^{\dagger}$}   & Time-bin & 20 km& 0.0063\\
Valivarthi et al., 2017  & BB84 & Time-bin & 80 km& 100\\
Liu et al., 2019  & BB84 & Time-bin & 160 km& 2.6\\
Wei et al., 2020  & Asymmetric MDI & Polarization & 180 km& 31\\
Zhou, et al., 2021  &\text{RFI}\tnote{$^{\dagger}$}  & Time-bin & 200 km & 1 \\
\hline %inserts single line
\end{tabular}
   \begin{tablenotes}
     \item[$^{\dagger}$] Reference-frame-independent.
   \end{tablenotes}
\end{threeparttable}
\label{table:nonlin2} % is used to refer this table in the text
\end{table}
%\vspace{-12pt}
%\footnotesize{$^{\dagger}$ Reference-frame-independent}\\

%Next table starts
\begin{table}[htb!]
\caption{Twin-field (TF) QKD implementations \cite{minder2019experimental,wang2019beating,liu2019experimentaltwin,zhong2019proof,fang2020implementation,liu2021field,wang2022twin}} % title of Table
\begin{threeparttable}[t]
\centering % used for centering table
%\fontsize{9}{11}\selectfont
\begin{tabular}{|p{2.3cm}| p{1.2cm}|  p{1.3cm}| p{0.9cm}| p{0.9cm}|} % centered columns (4 columns)
\hline %inserts double horizontal lines
Author, & Protocol & Encoding  & Channel & Key rate \\
Year    &         & scheme & length & bits/s \\
\hline % inserts single horizontal line
Minder et al., 2019  & TF & Phase  & 90.8 dB & 0.045  \\
Wang et al., 2019 & \text{SNS-}\text{TF}\tnote{$^{\dagger\dagger}$}  & Time-bin & 300 km & 2.01 K\\
Liu et al., 2019  & TF & Time-bin & 300 km & 39.2\\
Zhong et al., 2019  & TF & Phase & 55.1 dB & 25.6\\
Fang et al, 2020  & TF & Phase & 502 km & 0.118\\ % [1ex] adds vertical space
Liu, Hui, et al, 2021  &  \text{SNS-}\text{TF}\tnote{$^{\dagger\dagger}$} & Time-bin & 428 km & 3.36 \\
Wang, et al, 2021  & TF & Phase & 830 km & $~$0.01 \\
\hline %inserts single line
\end{tabular}
\begin{tablenotes}
     \item[$^{\dagger\dagger}$] Sending-or-not-sending twin-field.
   \end{tablenotes}
    \end{threeparttable}%
%\footnotesize{$^{\dagger\dagger}$ Sending-or-not-sending twin-field}
\label{table:nonlin3} % is used to refer this table in the text
\end{table}
%\vspace{-12pt}
%
%Next table starts
%Next table starts
\begin{table}[h!]
\caption{Continuous variable-QKD \cite{wang2017experimental,zhang2019continuous,zhang2020long}} %\cite{wang2017experimental, zhang2019continuous,zhang2020long}} % title of Table
\begin{threeparttable}[t]
\centering % used for centering table
%\fontsize{9}{11}\selectfont
\begin{tabular}{|p{2.2cm}| p{1.2cm}|  p{1.3cm}| p{0.9cm} |p{1.0cm}|} % centered columns (4 columns)
\hline %inserts double horizontal lines
Author, & Protocol & Encoding  & Channel & Key rate \\
Year    &         & scheme & length & bits/s \\
\hline % inserts single horizontal line
Wang et al., 2017  & CV & Gaussian modulation & 50 km & 700\\
Zhang et al., 2019  & CV & Gaussian modulation & 50 km & 5800\\
Zhang et al., 2020  & CV & Gaussian modulation & 202.8 km & 6.2\\
% [1ex] adds vertical space
\hline %inserts single line
\end{tabular}
\end{threeparttable}
\label{table:nonlin4} % is used to refer this table in the text
\end{table}
%\end{appendices}

\footnotesize{
\bibliographystyle{IEEEtran}
\bibliography{dps.bib}
}
\end{document}